\documentstyle[preprint,aps]{revtex}

\begin{document}


\def\beqar{\begin{eqnarray}}
\def\ra{\rightarrow}
\def\eeqar{\end{eqnarray}}
\def\be{\begin{eqnarray}}
\def\ee{\end{eqnarray}}
\def\beqast{\begin{eqnarray*}}
\def\eeqast{\end{eqnarray*}}
\def\be{\begin{enumerate}}
\def\ee{\end{enumerate}}
\def\lag{\langle}
\def\rag{\rangle}
\def\fnote#1#2{\begingroup\def\thefootnote{#1}\footnote{#2}
\addtocounter{footnote}{-1}\endgroup}
\def\beq{\begin{equation}}
\def\eeq{\end{equation}}

\draft
\preprint{
\begin{tabular}{r}
KAIST-TH 00/5
\\
hep-ph/0003309
\end{tabular}
}


\title{
$e^+ e^- \rightarrow {\tilde t}_1 {\tilde {t_1}}^* (H_1)$ in the MSSM
with explicit CP violations 
}

\author{
Saebyok Bae
\thanks{E-mail: dawn@muon.kaist.ac.kr}
}

\address{
Department of Physics, Korea Advanced Institute of Science and
Technology \\
Taejon 305-701, Korea \\
}

\maketitle

\ \\
\ \\
\begin{abstract}
We consider the effects of the CP-violating phases, {\it e.g.} 
$\arg(A_t)$ and $\arg(\mu)$ on the $e^+ e^-
\rightarrow {\tilde t}_1 {\tilde {t_1}}^* (H_1)$ processes.
The third generation squark trilinear terms give significant contributions to
the Higgs potential at the one-loop level. This results in the changes of 
the stop masses, the Higgs mass and the lighter 
stop-$\mbox{the lighter stop}^*$-the lightest Higgs coupling. We show the 
coupling and the loop effects on the processes. And we will discuss the 
determination method of soft parameters.
\end{abstract}

\tighten

\newpage

\section{Introduction}
The Standard Model (SM) can explain the CP violation of the $K$ system by 
means of the single KM phase in the CKM mixing matrix. And this paradigm 
will be tested in detail at B factories by measuring various CP violations 
in the $B$ system. The latter will constitute an important test for the SM in 
the CP violating sector, and may give a hint for new physics related with 
CP violations beyond the KM phase \cite{MASI} (which is strongly motivated
in order to properly understand baryogenesis). 
The best motivated candidates for new physics are models with supersymmetry
(SUSY), since SUSY can eliminate the quadratic divergences of the Higgs 
mass under grand unification theory (GUT).  This is 
one of fine tuning problems of the Standard Model, the so--called gauge 
hierarchy problem. Therefore it would be natural to consider the effects of 
CP violation in the context of the supersymmetric theories. The simplest 
extension of the Standard Model is the Minimal Supersymmetric Standard Model 
(MSSM), which we consider in the following. 
The CP-violating parameters in the MSSM contribute to the electric dipole 
moments (EDMs) of electron and neutron, or $\epsilon_K$ depending on 
their flavor structures, so that these $CP$ violating observables do constrain
the CP violating phases in the MSSM.  For example, the one loop contribution 
in the SUSY models to the  neutron EDM is schematically given by 
$d_n \sim 2 (100{\rm \,GeV}/{\tilde m})^2 \sin\phi \times 10^{-23} e \,
{\rm cm}$, where ${\tilde m}$ is the overall SUSY scale and $\phi$ is the 
typical CP violating phase \cite{GR etal}. The latest experimental bound is 
$|d_n| < 6.3 \times 10^{-26} e \,{\rm cm}$ (90\% C.L.) \cite{HA etal}. 
Therefore one can imagine various scenarios satisfying this tight experimental
constraint. 
In the usual scenario, one assumes that the CP phase $|\phi| \le 10^{-2}$ 
and the typical supersymmetric mass $\tilde{m} \le 1$ TeV. 
Or $\phi$ can be $O(1)$ and the scalar masses of the
first two generations are $O(1)$ TeV \cite{{MASI},{GR etal}}---this
is called the effective SUSY model. 
Recently, it was known that two independent CP violating phases in the minimal
supergravity GUT model \cite{{DU etal},{FAOL}} can be $O(1-10^{-1})$ 
with the typical supersymmetric masses $\tilde{m} \le 1$ TeV,  if some 
internal cancellations among various contributions occur \cite{IBNA}. In this
context of the internal cancellation mechanism of EDMs, two people \cite{DEMO}
analyzed the CP violating phase effects on $gg \rightarrow \Phi^0$ 
($\Phi^0=h^0, H^0, A^0$) for the parameter space where the Higgs mixing effects
\cite{PIWA} are negligible.

There are several important effects of the CP violating phases in the MSSM 
that have been studied recently. Firstly, the Higgs  
potential can be CP violating due to the one-loop effects of the third 
generation trilinear soft terms, although the tree level potential of the 
Higgs fields is CP invariant. So the lightest neutral Higgs mass can be 
larger or smaller than that of the CP conserving case (this can be 
phenomenologically important) \cite{PIWA}, and the branching ratios of Higgs 
bosons can be changed significantly \cite{CHLE}. Secondly, the large CP 
violating phases may be coincident with the cosmological upper bound of 
the relic density of neutralinos \cite{FAOL}, which are a candidate 
of dark matters in the R-parity invariant model. Thirdly, the
chargino pair-production has a strong dependence of $\arg(\mu)$, but it still 
can be used in determination of some SUSY parameters such as $\tan \beta$, 
gaugino mass $M_2$, $|\mu|$ and $\cos (\arg(\mu))$ \cite{CH etal}. Fourthly, 
the direct CP asymmetry $A_{\rm CP}^{b \rightarrow s \gamma}$ of 
$B \rightarrow X_s \gamma$ can be as large as $\pm$16\% \cite{BAKO} (which is 
much larger than the SM contribution $\sim 0.5$\% \cite{KANE}), if chargino 
and stops are light enough. Therefore this may be a good place to look for 
a new source of CP violation. Also the CP phases contribute to the CP 
violating $\epsilon_K$ parameter of the neutral Kaon \cite{GR etal}. 

In the MSSM, the lighter stop may be the lightest squark \cite{DREB}, because
the largeness of the top Yukawa coupling makes (1) the diagonal components of
stop $\rm{mass}^2$ matrix smaller than those of two other generation squarks
via renormalization group equations (RGEs) and (2) the off-diagonal components 
of stop $\rm{mass}^2$ matrix larger than those of other squarks 
\cite{{HI etal},{KONO}}.
Therefore, lighter stop may be produced relatively easily in the next
generation colliders. Furthermore, the stop has a large Yukawa coupling. 
This means that the associated Higgs productions with the stop-$\mbox{stop}^*$
pair could be sizable in the colliders\cite{DJ etal}. These subjects have been
considered by several groups in the MSSM without CP violations from the SUSY
sector. In the presence of CP violating phases in $A_t$ and $\mu$ parameters,
the masses and the mixing parameters of stops and Higgs bosons, and also the 
stop-$\mbox{stop}^*$-Higgs couplings would be modified. In this work, we 
consider these effects on the stop pair productions (with the lightest Higgs 
boson) at linear colliders.

This paper is organized as follows. In Sec.~II, we study the one-loop
effective potential of Higgs fields, the CP violating vacuum, 
the mixing of the different CP-property Higgs particles,
the stop $\rm{mass}^2$ matrix, and the stop-$\rm{stop}^*$-Higgs vertex. 
In Sec.~III, we will choose an appropriate MSSM scenario satisfying the 
electron/neutron EDM constraints. And we analyze the CP phase dependences and 
the magnitudes of the relative vacuum angle $\xi$, stop mass and the cross 
sections of the $e^+e^-\rightarrow {\tilde t}_1 {\tilde {t_1}}^* (H_1)$
processes. And we will discuss the determining method of the soft 
parameters in Sec.~IV, and we conclude in Sec.~V.

\section{The Higgs and stops sector in the MSSM}

\subsection{Higgs sector}

In the MSSM, it is well known that the Higgs F/D-term potentials are CP 
conserving at the tree level, and CP violating terms can reside in 
the soft SUSY breaking sector only. However, the squark-antisquark-Higgs 
trilinear soft terms can break CP, and they can generate {\it effective} 
CP violating terms in the Higgs effective potential via the squark loop 
corrections: namely, CP can be broken effectively in the Higgs sector. 
In terms of the Wilsonian action, the CP violating effective 
interactions occur when the higher frequency modes are integrated out in the 
CP-violating MSSM \cite{PESC}. But the Wilsonian action may not be good, 
since it can be unphysical when a massless particle participates \cite{SHVA}.
In Ref.~\cite{SHVA}, the electric charge of the scalar field 
in the Wilsonian action is dependent on the gauge fixing parameter $\zeta$ at
the two-loop level in the massless scalar QED. This is related to the infrared 
effect. And the field-integrating-out effective action 
(${\it e. g.}$ the four Fermi interaction with $W$ boson integrated out)
is not good for this case either, since the integrated-out field, 
the lighter stop can be lighter than the heaviest Higgs field which can be
included in the effective theory. Therefore, the 1PI effective action can 
be more suitable. The U(1)$_Y$ toy model \cite{PI297} can illustrate how the 
CP violating effective terms can be brought about by means of the 1PI 
effective action. 

The 1-loop corrected effective potential of the Higgs fields (we will follow 
the notations of Ref.~\cite{PIWA}, where many important calculations were
done) is given by
\beqar
\label{VHiggs}
{\cal V}_ {Higgs}^{eff}
&=&
\mu_{1}^2 \Phi_{1}^{\dagger}\Phi_{1}+\mu_{2}^2 \Phi_{2}^{\dagger}\Phi_{2}
+(m_{12}^2 \Phi_{1}^{\dagger} \Phi_{2} + h.c.)
\nonumber
\\
&&+
\lambda_{1}(\Phi_{1}^{\dagger}\Phi_{1})^2+\lambda_{2}(\Phi_{2}^{\dagger}
\Phi_{2})^2
+\lambda_{3}(\Phi_{1}^{\dagger} \Phi_{1})(\Phi_{2}^{\dagger} \Phi_{2})
+\lambda_{4}(\Phi_{1}^{\dagger} \Phi_{2})(\Phi_{2}^{\dagger} \Phi_{1})
\nonumber
\\
&&+
\lambda_{5}(\Phi_{1}^{\dagger}\Phi_{2})^2+\lambda_{5}^* (\Phi_{2}^{\dagger}
\Phi_{1})^2
+\lambda_{6}(\Phi_{1}^{\dagger} \Phi_{1})(\Phi_{1}^{\dagger} \Phi_{2})
+\lambda_{6}^*(\Phi_{1}^{\dagger} \Phi_{1})(\Phi_{2}^{\dagger} \Phi_{1})
\nonumber
\\
&&+
\lambda_{7}(\Phi_{2}^{\dagger} \Phi_{2})(\Phi_{1}^{\dagger} \Phi_{2})
+\lambda_{7}^*(\Phi_{2}^{\dagger} \Phi_{2})(\Phi_{2}^{\dagger} \Phi_{1})
~,
\eeqar
where $\lambda_i =0$ ($i=5,6,7$) at the tree level so that they are generated 
entirely by quantum corrections, and $\Phi_i$ $(i=1,2)$ are the scalar 
components of the Higgs superfields. The $\Phi_2 (\Phi_1)$ gives masses to 
the up-type (down-type) fermions. For $\tan\beta \sim O(1)$,
\beqar
\label{m12}
m_{12}^2 
&=&
B\mu + l \cdot h_t^2 A_t \mu~,
\nonumber
\\
\label{l}
l
&=&
\frac{1}{4 \cdot 16 \pi^2} 
\left(
\frac{m_{{\tilde t}_1}^2+m_{{\tilde t}_2}^2}{m_{{\tilde t}_1}^2-m_{{\tilde
t}_2}^2} \ln{\frac{m_{{\tilde t}_1}^2}{m_{{\tilde t}_2}^2}} - 2 
\right)~,
\eeqar
where $B$ is the soft parameter of the bilinear term, $h_t=\sqrt{2}
\, m_t(\bar{m_t}) / v \sin\beta$ and 
$m_{{\tilde t}_i}$ $(i=1, 2)$ are the masses of the lighter and 
heavier stops. The real parameter $l$ is $O(10^{-3})$ for appropriate parameter 
ranges, and $B \mu$ is set to be real by the field-redefinition or U(1)$_Q$ 
spurion transformation (this is our convention) \cite{DU etal}. 
Because the quantum correction is proportional to (Yukawa coupling)$^2$, the 
contribution of the 3rd generation squarks are larger than those of 1st/2nd 
generation squarks. Therefore, only stop contributions 
were written in Eq.~(\ref{m12}) at low $\tan\beta$, 
for which the sbottom contributions become
negligible. The parameter $B \mu$ denotes the mixing of two would-be CP-even 
Higgs bosons (scalar-scalar mixing), and $l \cdot h_t^2 A_t \mu$ represents 
the mixing of three CP-even and CP-odd Higgs bosons (scalar-pseudoscalar 
mixing). So $m_{12}^2$ plays an important role in the mixing of Higgs fields. 
If $|l \cdot h_t^2 A_t \mu| \ll |B \mu|$ and $\mu_1^2 \sim \mu_2^2 \sim 
|B \mu|$, we can expect that the scalar-pseudoscalar mixing is much smaller 
than the scalar-scalar mixing. Therefore, 
the mixing matrix $\cal O$ in Eq.~(\ref{Oij}) 
has a weak dependence of CP violating phases. In order to see the
scalar-pseudoscalar mixing more clearly, the authors of Ref.~\cite{PIWA}
used $|A_t|$, $|\mu|$ $\ge 1$ TeV, which may not be natural in the MSSM. 

The SU(2)$_L \times$ U(1)$_Y$ gauge symmetry is broken spontaneously into 
U(1)$_{\rm em}$ by the vacuum expectation values of two Higgs doublet fields 
\cite{PIWA}:
\beqar
\label{phi1}
\Phi_{1}
&=&
\left(
\begin{array}{c}
\phi_{1}^{+}
\\
(v_{1}+\phi_{1}+ia_{1})/\sqrt{2}
\end{array}
\right)~,
\nonumber
\\
\label{phi2}
\Phi_{2}
&=&
e^{i \xi}
\left(
\begin{array}{c}
\phi_{2}^{+}
\\
(v_{2}+\phi_{2}+ia_{2})/\sqrt{2}
\end{array}
\right)~,
\eeqar
where the VEVs $v_i$ are real, and $\langle \Omega|\{ \phi_{i}^{+},~ \phi_{i}, 
~a_{i} \}|\Omega \rangle=0$ $(i=1,2)$. 
The relative phase $\xi$ is determined from the minimum energy conditions of 
the Higgs potential \cite{PIWA}, which are nothing but 
$T_{\phi}={\partial{\cal V}_{Higgs}^{eff}}/{\partial \phi}=0$ 
at the vacuum $| \Omega \rangle$ ($\phi$ represents a scalar field).

Due to the CP violating
terms, {\it e.g.} $m_{12}^2 \Phi_{1}^{\dagger} \Phi_{2} + h.c.$, there exists
a scalar-pseudoscalar transition, which results in the mixing of the 
Higgs fields with different CP quantum numbers in the limit of CP invariant
Higgs sector.
So the $\rm{mass}^2$ matrix of the neutral Higgs particles \cite{PIWA} is
\beqar
\label{higgs mass}
{\cal M}^2_H
&=&
\left(
\begin{array}{cc}
{\cal M}^2_P  & {\cal M}^2_{PS} \\
{\cal M}^2_{SP} &  {\cal M}^2_S
\end{array}
\right)~,
\eeqar
where ${\cal M}^2_{SP}={{\cal M}^2_{PS}}^{T} \neq 0$.
Because the  symmetries are spontaneously  broken, there exists a 
{\it would-be}
Goldstone boson, $G^{0}$. From the Goldstone's theorem, $G^{0}$ should be
massless
for all the orders (from the tree level to all the loop levels). In the weak 
basis $(G^{0},~ a=-a_1 \sin \beta+a_2 \cos
\beta,~ \phi_{1},~ \phi_{2})$, ${{\cal M}_H^2}_{1j}=0$ $(j=2, 3, 4)$ 
\cite{PIWA}. We will define a new matrix ${\cal M}_{N}^2$ as
${{\cal M}_{N}^2}
_{ij}={{\cal M}_H^2}_{i+1,\, j+1}$ $(i,j=1,2,3)$.
${\cal M}_N^2$ is a real and symmetric matrix, so there exists a $3 \times 3$
orthogonal matrix ${\cal O}$ \cite{PIWA}, which satisfies
\beqar
\label{Oij}
{\cal O}^{T} {\cal M}^2_{N} {\cal O}= 
\mbox{diag} (M_{H_{3}}^2, \, M_{H_{2}}^2, \, M_{H_{1}}^2 )~,
\eeqar
where $M_{H_{3}} \geq M_{H_{2}} \geq M_{H_{1}}$. The corresponding mass
eigenstates, $H_{i}$ $(i=1,2,3) \,$, are
\beqar
\label{s-psmixing}
\left(
\begin{array}{c}
H_3\\
H_2\\
H_1
\end{array}
\right)
=
{\cal O}^{T}
\left(
\begin{array}{c}
a\\
\phi_1\\
\phi_2
\end{array}
\right)~.
\eeqar

\subsection{Stop sector}

If we substitute the $\Phi_i$'s of Eqs.~(\ref{phi2}) 
into the MSSM Lagrangian, the stop $\rm{mass}^2$ matrix 
${\cal M}_{\tilde t}^2$ is in the 
\beqar
\label{stop mass}
{\cal L}_{mass}^{eff} &=&
- \, ( \tilde{t}^*_L ~ \tilde{t}^*_R )
{\cal M}^2_{\tilde{t}}
\left(
\begin{array}{c}
\tilde{t}_L
\\
\tilde{t}_R
\end{array}
\right)
\nonumber
\\
&=&
- \, ( \tilde{t}^*_L ~ \tilde{t}^*_R )
\left(
\begin{array}{cc}
m_{\tilde{t}L}^2
&m_{\tilde{t}LR}^2
\\
m_{\tilde{t}LR}^{2*}
&m_{\tilde{t}R}^2
\end{array}
\right)
\left(
\begin{array}{c}
\tilde{t}_L
\\
\tilde{t}_R
\end{array}
\right)
~,
\eeqar
where
\beqar
\label{mL}
m_{\tilde{t}L}^2 &=&
M_{\tilde{t}_L}^2 + m_t^2 + M_Z^2
\cos{2\beta}\left(\frac{1}{2}-\frac{2}{3}\sin^2\theta_W\right)~,
\nonumber
\\
\label{mR}
m_{\tilde{t}R}^2 &=&
M_{\tilde{t}_R}^2 + m_t^2 + M_Z^2
\cos{2\beta}\cdot\frac{2}{3}\sin^2\theta_W~,
\\
\label{mLR}
m_{\tilde{t}LR}^2 &=&
m_t \, (A_t^* e^{-i \xi}-\mu \cot\beta)~.
\nonumber
\eeqar
Although $\xi$ is $0$ or $\pi$ (mod $2\pi$) by the field redefinition
at the tree level, it cannot be no longer the case when one includes 
the 1-loop effects \cite{PI373}, which is determined from the minimum energy 
conditions $T_{\phi}={\partial{\cal V}_{Higgs}^{eff}}/{\partial \phi}=0$ 
\cite{PIWA}. Therefore, the stop mixing angle $\theta_{\tilde t}$ 
\cite{IBNA} is changed via
\beq
\label{theta t}
\tan{2\theta_{\tilde t}} =
\frac{2|m_{\tilde{t}LR}^2|}
{m_{\tilde{t}L}^2-m_{\tilde{t}R}^2}~.
\eeq
The relations between the mass and the weak eigenstates of stops are given by
\beqar
\label{mw transf1}
{\tilde{t_1}} &=&
\tilde{t}_L \cos \theta_{\tilde t} +\tilde{t}_R \,e^{-i\beta_{\tilde t}} 
\sin \theta_{\tilde t}~,
\nonumber
\\
\label{mw transf2}
{\tilde{t}_2} &=&
-\tilde{t}_L \, e^{i\beta_{\tilde t}} \sin \theta_{\tilde t} 
+\tilde{t}_R \cos\theta_{\tilde t}~,
\eeqar
where $\beta_{\tilde t}=-\arg (m_{\tilde{t} LR}^2)$.
The mass eigenvalues of two stops are 
\beq
\label{mt1,2}
m^2_{{{\tilde t}_1},{{\tilde t}_2}} 
=
\frac{m_{{\tilde t}L}^2+m_{{\tilde t}R}^2
\mp \sqrt{(m_{{\tilde t}L}^2-m_{{\tilde
t}R}^2)^2+4|m_{{\tilde t}LR}^2|^2}}{2}~.
\eeq
Note that $m_{{{\tilde t}_1},{{\tilde t}_2}}^2$ is dependent on the CP 
violating phases, $\arg( A_t )$ and $\arg( \mu )$  due to 
$m_{{\tilde t}LR}^2$ in Eqs.~(\ref{mR}).

\subsection{Stop-$\mbox{stop}^*$-the lightest Higgs vertex}

In the presence of CP violations in the Higgs sector, the 
stop-$\mbox{stop}^*$-the lightest Higgs interaction is also modified. 
Defining the relevant interaction Lagrangian as 
\beq
\label{LttH}
{\cal L}_{{\tilde t}_1 {\tilde {t_1}}^* H_1}^{eff}
= 
- V_{{\tilde t}_1 {\tilde {t_1}}^* H_1}^{eff} \cdot
{\tilde t}_1 {\tilde {t_1}}^* H_1~,
\eeq
the coupling $ V_{{\tilde t}_1 {\tilde {t_1}}^* H_1}^{eff}$  is given by 
\beqar
\label{VttH}
V_{{\tilde t}_1 {\tilde {t_1}}^* H_1}^{eff}
&=&
-\frac{g_W M_W}{2 \cos^2 \theta_W}
({\cal O}_{33} \sin \beta -{\cal O}_{23}\cos \beta)
\left[
\left(\frac{1}{2}-\frac{2}{3}\sin^2\theta_W \right)\cos^2\theta_{\tilde t}
+\frac{2}{3}\sin^2\theta_W\sin^2\theta_{\tilde t}
\right]
\nonumber
\\
&&
+\frac{g_W m_t}{2 M_W \sin\beta}
{\bf Re} \left[
A_t e^{i(\xi - \beta_{\tilde t})}({\cal O}_{33}+i{\cal O}_{13} \cos \beta)
-\mu \, e^{i \beta_{\tilde t}}({\cal O}_{23}+i{\cal O}_{13}
\cos \beta)
\right]
\nonumber
\\
&&
+\frac{g_W m_t^2}{M_W \sin\beta}{\cal O}_{33}~.
\eeqar
The $g_W$ is the SU(2)$_L$ gauge coupling constant and the ${\cal O}_{i3}$ 
are determined from ${\cal M}_N^2$ \cite{PIWA}. Since ${\cal M}_N^2$ 
depends on the CP violating phases, so do the ${\cal O}_{i3}$'s ($i=1,2,3$).
Note that at the tree level (without CP violations in the Higgs sector), 
$H_1=h$, $H_2=H$, $\xi=0$ and the scalar-pseudoscalar mixing does not exist.
Therefore, we recover the usual expressions, 
\beqar
\label{O->}
{\cal O}_{13}
&=&
0~,
\nonumber
\\
{\cal O}_{23}
&=&
-\sin \alpha~,
\nonumber
\\ 
{\cal O}_{33}
&=&
+\cos \alpha~,
\nonumber
\eeqar
where $\alpha$ is the scalar Higgs mixing angle at the tree level. Due to 
the 1-loop corrections, the tree level parameters are deformed,
even inducing the mixing among Higgs bosons with different CP properties.

\section{
The Analyses of \mbox{$ e^+ e^- \rightarrow \tilde{t}_1 \tilde{t_1}^*
(H_1)$}}

\subsection{EDM constraints }

Before we discuss the CP violating phase dependences of the stop masses and 
their production cross sections at next linear colliders (NLC's), 
we have to consider
the neutron/electron  EDM constraints on the CP violating phases, 
$\arg(A_t)$ and $\arg(\mu)$.  In the previous section, $\arg(\mu)$ was 
transferred into $\arg(A_t)$ in the form of $\arg(A_t)+\arg(\mu)$ from  
Eqs.~(\ref{m12}) and (\ref{mR}). So we can take $\mu$ to be real, and vary 
$\arg(A_t)$ from $0$ to $2\pi$ in the following. 
In the minimal supergravity model, the phases $\arg(A_{t0})$ and $\arg(\mu_0)$ 
can be $O(1-10^{-1})$ without conflict with neutron/electron EDM 
constraints \cite{IBNA}. The $A_{t0}$ and $\mu_0$ are the values of 
$A_t$ and $\mu$ at the GUT scale. But these two phases are rather correlated,
namely having different signs, and $|\arg(A_{t0})+\arg(\mu_0)|$
is typically smaller than $O(\pi/9)$ for low $\tan\beta$ and the soft mass 
scale $\gtrsim 500$ GeV \cite{IBNA}. Of course, we should consider the 
renormalization group behaviors of $\arg(A_t)$ and $\arg(\mu)$ down to 
the electroweak (EW) scale. 
In the ordinary phase convention of the minimal supergravity GUT
(only $A$ and $\mu$ terms have CP violating phases),
if an appropriate universality at the GUT scale is assumed, 
${\bf{Re}}(A_{t{\rm EW}}) > {\bf{Re}}(A_{t0})$ and ${\bf{Im}}(A_{t{\rm
EW}}) < {\bf{Im}}(A_{t0})$ from the typical RGEs and 
the numerical values of Ref.~\cite{HI etal}, where $A_{t \rm{EW}}$ is the
value of $A_t$ at the EW scale. This means that the
$\arg(A_{t\rm{EW}}) < \arg(A_{t0})$ (about ten times smaller), which is 
different from the $\mu$ case, that is, $\arg(\mu_{\rm EW})=\arg(\mu_0)$ since  
${\bf Re}(\mu)$ and ${\bf Im}(\mu)$ have the same RGE, $\dot{y}(t)=a(t)y(t)$
\cite{HI etal}. So we can conclude that $|\arg(A_{t})+\arg(\mu)|$ decreases
from the GUT scale to the EW scale via RGEs.  
Therefore the phase effects on stop pair productions at NLC's would not be 
that prominent in this case, and thus is not proper for our analyses. 
This leads us to consider other scenarios with large CP phases.
We will choose a slightly non-universal scenario for the trilinear couplings 
$A_f$ \cite{{PIWA},{KE etal}}, the underlying assumptions of which are
\cite{KE etal} 
\begin{itemize}
\item $|\arg(\mu)| \lesssim 10^{-2}$ from cosmological reasonings 
\item $|A_e|, |A_{u,c}|, |A_{d,s}| \lesssim 10^{-3} |\mu|$
\item $A_t=A_b=A_{\tau}$, where the only large CP-violating phase 
is contained  
\item gluino mass $m_{\tilde g} \gtrsim 400$ GeV 
\end{itemize}
The first two assumptions are chosen in order to to suppress the one-loop EDM 
contributions  owing to $d_f^{\rm 1-loop} \propto 
\{ {\rm {\bf Im}}(\mu), {\rm {\bf Im}}(A_f)\}$ 
for electron and light quarks \cite{KE etal}. The two-loop contributions 
can be much smaller than the
experimental bounds if $|\mu|$, ${\tilde m}_2 \gtrsim$ 100 GeV \cite{KE etal}.
And the neutron EDM has a significant three-loop \cite{KE etal} contribution 
$\frac{e \Lambda_{\chi}}{4\pi} d^G(\Lambda_{\chi})$ from Weinberg's three gluon
operator ${\cal O}_{\rm 3-gluon}=-\frac{1}{6}d^G f_{abc}G^a_{\mu \rho}
G^{b \rho}_{\nu} G^c_{\lambda \sigma} \epsilon^{\mu \nu \lambda \sigma}$,
which comes from two-loop diagrams with top, stop and gluino internal lines 
\cite{{BR etal},{IBNA2}}. The $\Lambda_{\chi}$ is the chiral symmetry
breaking scale. It has been shown that the three-loop
contribution $\propto 1/m_{\tilde g}^3$ are much smaller than 
the experimental bound of neutron EDM if 
gluino mass $m_{\tilde g}$ is larger than 400 GeV \cite{{BR etal},{IBNA2}}.
So the 3-loop contribution to the EDM can be sufficiently 
smaller than the experimental bound by the fourth assumption. 
In Ref.~\cite{KE etal}, large $\tan \beta$ scenarios
($40 \lesssim \tan \beta \lesssim 60$) with $|\mu|, |A_t| > 500$ GeV, $M_a \leq
500$ GeV and large CP phase are excluded, but low $\tan \beta$ scenarios
($\tan \beta \lesssim 20$) can be possible. We will consider 
the parameter ranges, which are not ruled out by the EDM
constraints, {\it i.e.} $\tan \beta = 2 (\ll 20$), $|\mu| = 500,~ 1000$ GeV
and $B=30$ GeV, for which the  would-be CP-odd Higgs boson has its mass 
$M_a \approx \sqrt{|2 {\bf Re}(m_{12}^2 e^{i \xi})/\sin{2\beta}|}\approx 194, 
\,274$ GeV, evading the EDM constraints by the authors of Ref. \cite{KE etal}.
Also we choose $|A_t|=|\mu\cot\beta|$ in order to maximize the 
effects of the $A_t$ phase in the stop mass matrix.

\subsection{Vacuum angle $\xi$}

\begin{figure}
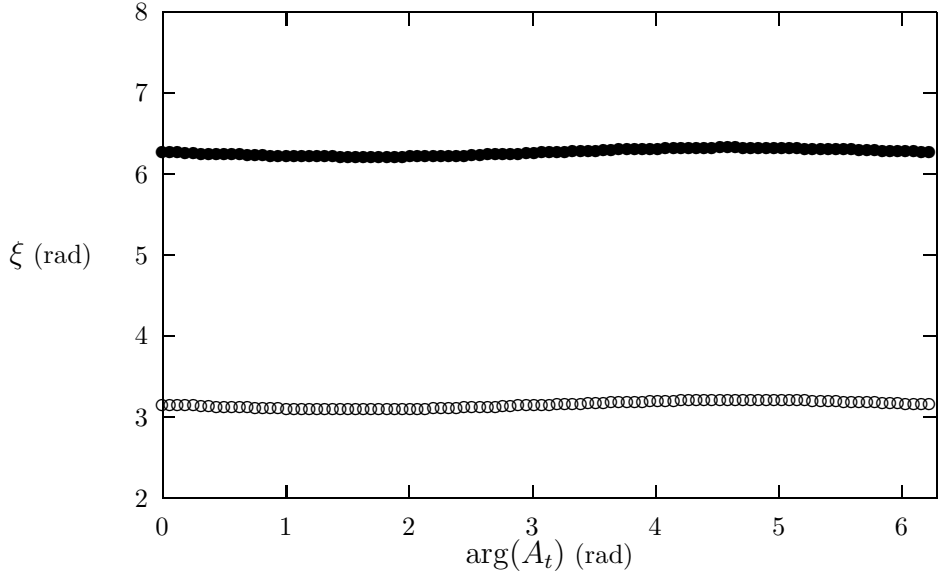

\begin{center}
\setlength{\unitlength}{0.240900pt}
\ifx\plotpoint\undefined\newsavebox{\plotpoint}\fi

\end{center}
\caption{\label{dxi12} The variations of $\xi$ with respect to
$\arg(A_t)$ ($\mu=500$ GeV, $|A_t|=250$ GeV, $M_a \approx 194$ GeV  
and $\tan \beta =2$ ) 
The $\bullet$
denotes the point which corresponds to $\xi(\arg(A_t)=0)=2 \pi$ (or 0)
at the tree 
level and the $\circ$ denotes  the point which corresponds to
$\xi(\arg(A_t)=0)=\pi$ at the tree level.}
\end{figure}

\begin{figure}
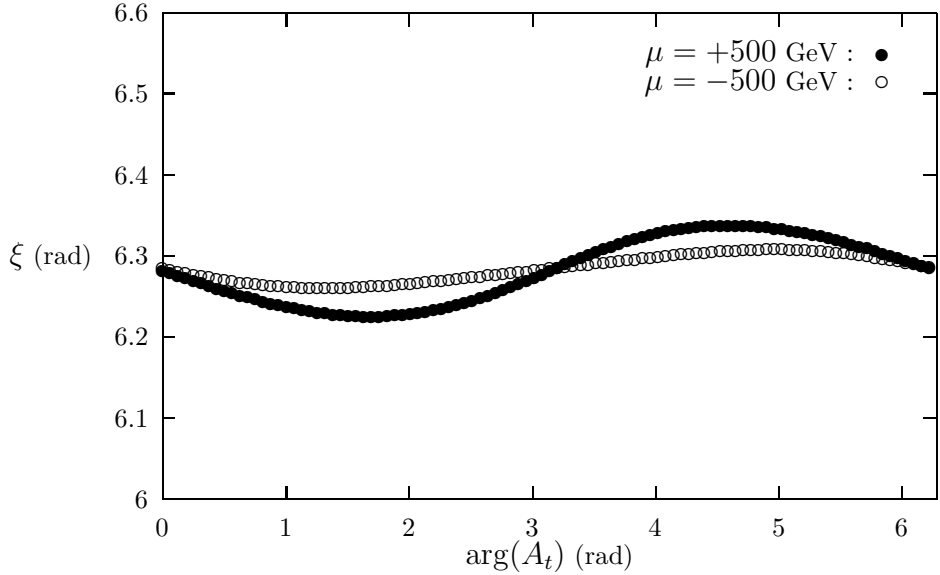

\begin{center}
\setlength{\unitlength}{0.240900pt}
\ifx\plotpoint\undefined\newsavebox{\plotpoint}\fi
%
\end{center}
\caption{\label{xi+-} The variations of $\xi$ 
with respect to the sign of $\mu$ (or $\arg(\mu)=0, \pi$)
($\mu=\pm 500$ GeV, $|A_t|=250$ GeV, $\tan\beta=2$ and $M_{\rm SUSY}=500$ 
GeV \,)}
\end{figure}

\begin{figure}
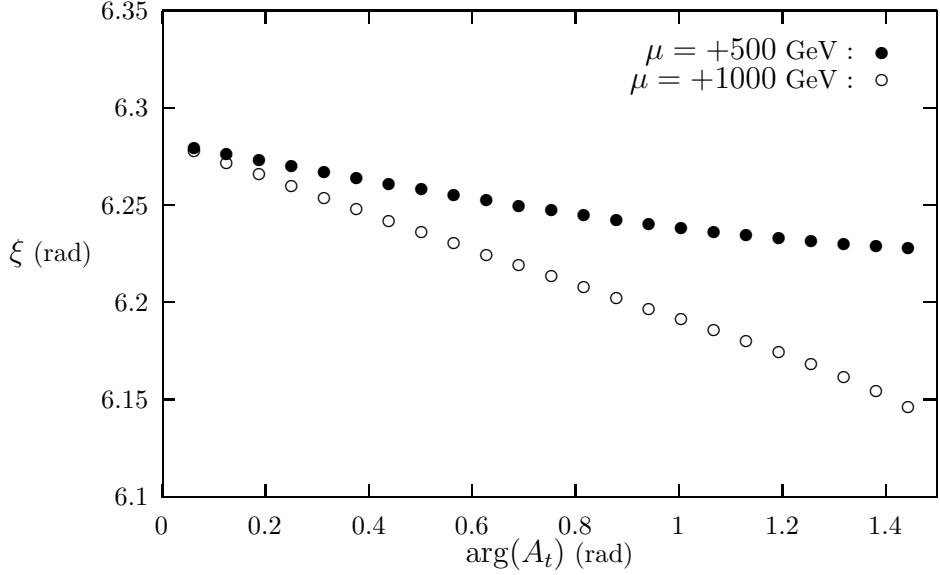

\begin{center}
\setlength{\unitlength}{0.240900pt}
\ifx\plotpoint\undefined\newsavebox{\plotpoint}\fi
%
\end{center}
\caption{\label{xm5-10} The variations of $\xi$'s 
with respect to the magnitude of $\mu$
($|A_t|=|\mu \cot\beta|$, $\tan\beta=2$ and $M_{\rm SUSY}=500$ GeV\,) 
Note that the range 
of $\arg(A_t)$ is [0, 1.5] approximately, because the lightest Higgs mass
$M_{H_1}$ is imaginary elsewhere owing to the large $|\mu|$.}
\end{figure}

We analyze the relative phase $\xi$ of two Higgs VEVs 
in Eqs.~(\ref{phi2}). The $\xi$ is a
solution of the condition of a vanishing non-trivial tadpole parameter
(see Ref.~\cite{PIWA})
\beqar
\label{Ta}
T_a &=& \frac{\partial {\cal{V}}_{Higgs}^{eff}}{\partial a}
= -v
\left[\,
{\bf Im}(m_{12}^2 e^{i \xi})
+{\bf Im}({\lambda}_5 e^{2 i \xi}) v^2 \sin\beta \cos\beta
+\frac{1}{2}{\bf Im}({\lambda}_6 e^{i \xi}) v^2 \cos^2\beta
\right.
\nonumber
\\
&&~~~~~~~~~~~~~~~~~~~~
\left.
+{\bf Im}({\lambda}_7 e^{i \xi}) v^2 \sin^2\beta
\,\right] = 0.
\eeqar
In Fig.~\ref{dxi12}, we show $\xi$ as a function of $\arg ( A_t )$ for 
$\mu = +500$ GeV. $\xi$ has two sets of values around $0$ and $\pi$ 
(mod $2\pi$).  If $\xi$ is neither $0$ nor $\pi$ (mod $2\pi$), CP is 
spontaneously broken by the vacuum due to the quantum corrections to the 
Higgs potential coming from the stop loop. However the effect on $\xi$ is not 
that prominent since $\arg(m_{12}^2) \approx 0$ and $|m_{12}^2| \gg
|\lambda_{5-7}v^2|$ for our parameter region in Eq.~(\ref{Ta}). 
In Fig.~\ref{xi+-}, we show $\xi$ for the same set of parameters except that
the sign of $\mu$ takes both positive and negative. The maximal deviations of
$\xi$ from $2\pi$ is again less than $O(2\pi/120)$ ({\it i.e.} has 
a weak phase dependence) and varies with the sign of $\mu$. 
We calculated the results for a real $\mu$ with two different signs 
({\it i.e.} $\arg(\mu)=0~{\rm or}~\pi$) in order to see the effects of 
changing $\arg(\mu)$, even though such a negative sign of $\mu$ can not 
satisfy the first assumption of the
slightly non-universal scenario, strictly speaking.  If the magnitude of 
$\mu$ gets larger (about 1 TeV), the $\xi$ has a smaller range of $\arg(A_t)$ 
as shown in Fig.~\ref{xm5-10}, since the lightest Higgs mass 
$M_{H_1}$ gets smaller and finally becomes imaginary elsewhere. This latter 
phenomenon was observed in Figs.~2(a) and 6(a) of Ref.~\cite{PIWA}. 
Also the sizes of $\xi-\pi({\rm or}~ 2\pi)$ become larger than those of the 
smaller $\mu$ case, as shown in Fig.~\ref{xm5-10}. 

In addition, $\xi$ appears in the mass matrices of charginos and neutralinos. 
(see Refs.~\cite{{HI etal},{MA}}.) But the sizes of $\xi-\pi({\rm and}~ 2\pi)$ 
are very small, so they can not have significant 1-loop effects 
on charginos and neutralinos. $\xi$ may also appear in the strong CP 
parameter ${\bar \theta}= \theta_{\rm QCD}+\arg({\rm det}(M_{\rm quark}))
=\theta_{\rm QCD}+\arg({\rm det}(M_u M_d))$ \cite{PE} 
through the Standard Model Yukawa terms since the masses of the up-type and
down-type quarks are proportional to the VEVs of Higgs fields in 
Eqs.~(\ref{phi2}). The $M_{\rm quark}$ is an $N_f \times N_f$ matrix, 
where $N_f$ is the number of flavors, and equal to 6 in the SM.

\subsection{$e^+ e^- \rightarrow {\tilde{t_1}} {\tilde{t_1}}^*$ process}

\begin{figure}
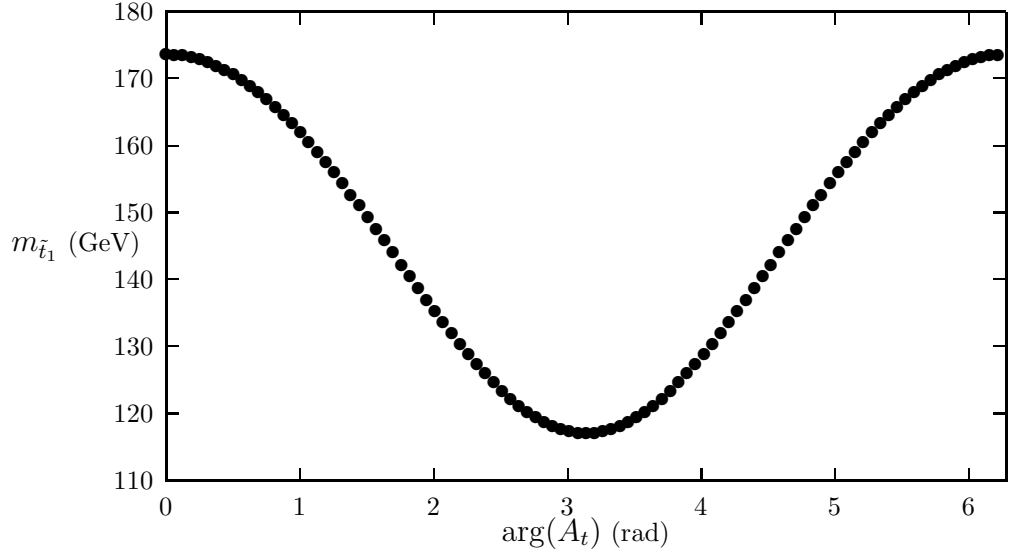

\begin{center}
\setlength{\unitlength}{0.240900pt}
\ifx\plotpoint\undefined\newsavebox{\plotpoint}\fi
%
\end{center}
\caption{\label{mt1} The variations of the lightest stop mass 
with respect to $\arg(A_t)$ ($\mu=500$ GeV, $|A_t|=250$ GeV,
$\tan\beta=2$ and 
$M_{\rm SUSY}=500$ GeV\,)}
\end{figure}

Let us first consider a process, 
$e^+ e^- \rightarrow {\tilde{t_1}} {\tilde{t_1}}^*$ at NLC's with 
$\sqrt{s}=500$ GeV. As before, we take $\mu=500$ GeV, $\tan \beta =2$, 
$M_{\rm SUSY}=500$ GeV, and finally $|A_t|=|\mu \cot\beta|$ in order to 
maximize the effects of $\arg(A_t)$.  Otherwise, the effects of 
$\arg ( A_t )$ relative to $\arg \mu $ will be negligible: see  
Eqs.~(\ref{mR}), (\ref{theta t}) and (\ref{mt1,2}).  Since $\sqrt s \gg M_Z$ at 
NLC's, we neglect $M_Z$ in the $Z$ boson propagator. 
The amplitude for this process 
with a left-handed initial electron is
\beq
\label{Mtt}
{\cal M}_{{\tilde {t_1}} {\tilde {t_1}}^*}^{L} 
=
\frac{ g_W^2 \sin^2 \theta_W}{s}
(V_1+V_2)  \, \bar{v}_L(p') \, \gamma^\mu
\, u_L(p)(k'_\mu-k_\mu)~, 
\eeq
where $p$ and $p'$ are 4-momenta of $\,e^-$ and $e^+$
respectively, $s=(p+p')^2$ and 
\beqar
\label{V1}
V_1 &=&
\frac{2}{3}-\frac{1}{\sin^2\theta_W \cos^2\theta_W}
\left(\sin^2\theta_W-\frac{1}{4} 
\right)
\left(\frac{1}{2}\cos^2\theta_{\tilde t}
-\frac{2}{3}\sin^2\theta_W
\right)~,
\\
\label{V2}
V_2 &=&
\frac{1}{4 \sin^2\theta_W \cos^2\theta_W}
\left(\frac{1}{2}\cos^2\theta_{\tilde t}
-\frac{2}{3}\sin^2\theta_W
\right)~.
\eeqar
The amplitude with a right-handed electron ${\cal M}_{{\tilde t}_1
{\tilde {t_1}}^*}^{R}$ is obtained by replacement $u_L,~ v_L \rightarrow 
u_R,~ v_R$ and $V_1+V_2 \rightarrow V_1-V_2$ in Eq.~(\ref{Mtt}). 
We can measure the $\cos^2\theta_{\tilde t}$ from 
$\sigma_L/\sigma_R$ \cite{BE etal}, where $\sigma_i$ $(i=L, R)$ are  
the cross sections corresponding to the polarizations of the electrons. In
Ref.~\cite{BE etal}, this process was considered at the tree level in the 
MSSM with real $\mu$ and $A_t$. In the presence of $A_t$ phase, the cross 
sections could be altered significantly if $| A_t | \sim | \mu \cot\beta |$, 
since the physical stop mass and the mixing angle $\cos^2\theta_{\tilde t}$ 
($V_1$ and $V_2$) depend strongly on the phase $\arg(A_t)$ through 
Eqs.~(\ref{theta t}) and (\ref{mt1,2}) due to the fact that 
$|m_{{\tilde t}LR}^2|$ in Eqs.~(\ref{mR}) can vary from $0$ to $2 m_t |A_t|$. 
Because the phase dependence of $m_{{\tilde t}_1}$ becomes prominent for our
parameter region from Fig.~\ref{mt1}, the cross sections can depend on 
the phase strongly, since  $(1-4 m_{{\tilde t}_1}^2/s)^{3/2}$ of the cross 
sections has a much stronger phase dependence through the stop mass than 
the coupling $(V_1^2+V_2^2)$.  
However, the 1-loop correction to the Higgs potential does not change the 
cross sections very much, which is not hard to understand. The phase 
dependences come from only $A_t^*$ and $e^{i\xi}$ in Eq.~(\ref{mR}). 
The 1-loop corrections hardly change $\xi$ which has a very small amplitude, 
so the unique phase $\xi$ that was generated by quantum corrections can be 
safely neglected. And, there is no significant change of the cross section 
as shown in Fig.~\ref{dstpm5}. 

The total cross section for $e^+ e^-  \rightarrow \tilde{t}_1 {\tilde{t_1}}^*$
at $\sqrt{s} = 500$ GeV is greater than 40 ${\rm fb}$ (see Fig.~\ref{dstpm5}).
Therefore, if a high integrated luminosity $\int {\cal{L}} dt = 500 ~
{\rm fb^{-1}}$ could be achieved at $\sqrt{s}=500$ GeV as suggested by TESLA 
working group, we would expect more than about 20000 events for a year.
This should be enough for studying the lightest stop mass, the cross section
and the main decay modes of the lightest stop. The branching ratios of the stop
in the CP conserving MSSM can be found in Ref.~\cite{BO etal}.

Since physical quantities such as stop masses and their production cross 
section depend strongly on $\arg(A_t)$, it is important to determine 
$\arg(A_t)$ experimentally. 
We will discuss this issue in Sec.~IV.  

\subsection{$e^+ e^- \rightarrow {\tilde{t_1}}{\tilde{t_1}}^* H_1$ process}

We analyze the process $e^+ e^- \rightarrow {\tilde{t_1}} 
{\tilde{t_1}}^* H_1$, where $\sqrt{s}=500$ GeV,
$\mu=500,~1000$ GeV, $\tan \beta =2$, $M_{\rm SUSY}=500$ GeV 
and again $|A_t|=|\mu \cot\beta|$.
At the tree level, the most significant diagrams have the ${\tilde {t_1}} 
{\tilde {t_1}}^* H_1$ vertex alone and other diagrams have
${\tilde {t_2}} {\tilde {t_1}}^{(*)} H_1$ and 
$Z^0 {\tilde {t_1}} {\tilde {t_1}}^*$ vertices
\cite{BE etal}. The 1-loop corrected amplitude 
with a left-handed electron is
\beqar
\label{MttH}
{\cal M}_{{\tilde {t_1}} {\tilde {t_1}}^* H_1}^{L}
&=&
- \frac{g_W^2 \sin^2\theta_W \cdot
V_{{\tilde {t_1}} {\tilde {t_1}}^* H_1}^{eff}}{s} 
\, (V_1+V_2) \,
\bar{v}_L(p') \, \gamma^\mu \,  u_L(p)
\nonumber
\\
&&~~~~~
\times
\left(
\frac{q'_\mu -k_\mu}{q'^2-m_{{\tilde {t_1}}}^2}
+\frac{k'_\mu - q''_\mu}{q''^2-m_{{\tilde {t_1}}}^2}
\right) ~,
\eeqar
where $q'=p+p'-k$, $q''=p+p'-k'$, $\,k$ is
the 4-momentum of ${\tilde {t_1}}$, 
and $\,k'$ is the 4-momentum of ${\tilde {t_1}^*}$.
The amplitude with a right-handed electron ${\cal M}_{{\tilde {t_1}} {\tilde
{t_1}}^* H_1}^{R}$
is obtained by replacement 
$u_L,~ v_L \rightarrow u_R,~ v_R$ and
$V_1+V_2 \rightarrow V_1-V_2$ in Eq. (\ref{MttH}).
The cross sections depend on the CP violating phase because 
$V_{{\tilde {t_1}} {\tilde {t_1}}^* H_1}^{\, eff}$ and 
$\cos^2\theta_{\tilde t}$ are varying
with the phase via Eqs. (\ref{higgs mass}), (\ref{Oij}), (\ref{mR}) 
and (\ref{theta t}). 

As noticed in the previous subsection, we can neglect the 1-loop effect coming
through $\xi$ due to its tiny variations around the tree level values. 
However, the quantum corrections can affect the 
${{\cal O}_{i3}}$ ($i=1,2,3$) and the Higgs mass $M_{H_1}$ significantly,
in general. But in Sec.~II, it was argued that the mixing matrix 
$\cal O$ can have a weak phase dependence in some region, and our parameters 
are exactly in that region. Therefore, only the lightest Higgs mass $M_{H_1}$ 
is much affected by the quantum corrections ($M_{H_1}$ is independent of
$\arg(A_t)$ at the tree level). This results in a sizable change of the cross
section for $e^+ e^- \rightarrow {\tilde{t_1}}{\tilde{t_1}}^* H_1$ through  
the strong dependence of the Higgs mass on the $A_t$ phase.
The numerical results are shown in Fig.~\ref{dcm5} (in filled circles)
along with the tree level results (in open circles).
Also the possible range for the cross section becomes much wider once we 
include the one-loop corrections to the Higgs potential. The symmetry of the
cross section about $\arg(A_t)=\pi$ in Fig.~\ref{dcm5} is represented by the
symmetry under $\arg(A_t) \rightarrow 2\pi-\arg(A_t)$, $i.e.$ $\cos(\arg(A_t))
\rightarrow \cos(\arg(A_t))$ and $\sin(\arg(A_t)) \rightarrow -\sin(\arg(A_t))$.
The Higgs masses \cite{PIWA} are 
\beq
\label{Mh}
M_{H_i}^2 = 
-\frac{1}{3}\, r + 2 \left( -\frac{p^3}{27} \right)^{1/6} \cos\left(\frac
{\phi}{3}+\delta_i \right)~,
\eeq
where 
\beqar
r &=& - {\rm tr} ({\cal M}_N^2)~,
\nonumber
\\
s &=& \left[ {\rm tr}^2({\cal M}_N^2)-{\rm tr}({\cal M}_N^4) 
\right]/2 ~,
\nonumber
\\
t &=& -{\rm det}({\cal M}_N^2)~,
\nonumber
\\
p &=& (3s-r^2)/3~,
\\
q &=& 2r^3/27-rs/3+t~,
\nonumber
\nonumber
\\
\phi &=& \cos^{-1} \left(
q/2\sqrt{-p^3/27} \right)~,
\nonumber
\\
\delta_i &=& 0, \pm 2\pi/3~ (i=1,2,3)~.
\nonumber
\eeqar
Note that ${\cal M}_N^2$ is of form $f(|B|, \xi, |A_t|, \arg(A_t), |A_b|,
\arg(A_b), M_{\rm SUSY}^2)$ without $\mu_i^2$ ($i=1,2$) by using  
two minimum energy conditions. Since the stop mass $m_{\tilde{t_1}}$ and 
the Higgs mass $M_{H_1}$ are symmetric under $\arg(A_t) \rightarrow 
2\pi-\arg(A_t)$ \cite{PIWA}, the cross section has that symmetry. 
And the cross sections have a minimum  near $\arg(A_t)=\pi$ in Fig.~\ref{dcm5}
because the coupling $(V_{{\tilde {t_1}} {\tilde {t_1}}^* H_1}^{eff })^2 
(V_1^2+V_2^2)$ has a minimum at $\arg(A_t)=\pi$ and its phase dependence 
is stronger than the dependence through the masses of stop and Higgs boson 
unlike the $e^+ e^- \rightarrow {\tilde{t_1}}{\tilde{t_1}}^*$ process. 
For larger $\mu$ (about 1 TeV), the qualitative features remain the same as 
the lower $\mu$ case, as shown in Fig.~\ref{dcm10}. But the larger $|\mu|$ 
brings about the region of $\arg(A_t)$ where the lightest Higgs mass 
$M_{H_1}$ becomes imaginary ({\it i.e.} unphysical) as discussed before.

And the 1-loop cross sections are sufficiently enhanced
from the tree level results. This can be a good news in the light of
experiment. For the integrated luminosity, $\int{\cal{L}} dt= 500$ 
${\rm fb^{-1}}$ at $\sqrt{s}=500$ GeV, one can have more than 250 events for 
a year, since the cross section is larger than 0.5 ${\rm fb}$ from Fig. 
\ref{dcm5}.  The study of the decay modes and the branching ratios of Higgs 
bosons in the presence of $\mu$ and $A_t$ phases can be found in 
Ref.~\cite{CHLE}.

\section{The determinations of the soft parameters}

In the previous section, we observed important physical quantities (stop 
mass and cross sections) have strong phase dependences. Therefore, it is
important to determine the value of the CP violating phase $\arg(A_t)$. 
In order to fix $n$ unknown parameters, we need $n$ independent equations 
involving these unknown parameters. The equations should be, of course, about 
physical observables like masses, cross sections and so on. Therefore, if we 
have enough independent physical quantities from experiments, the soft 
parameters may be determined in principle. For example, it is well known 
that $\tan\beta$, $|\mu|$ and $\arg(\mu)$ can be determined from $e^+ e^- 
\rightarrow \tilde{\chi_i}^+ \tilde{\chi_j}^-$ with 
$i,j = 1,2$ \cite{CH etal}. This was 
possible because charginos are spin-half particles, so that there are many 
physical observables with different combinations of chargino helicities 
\cite{CH etal}. Therefore, we will assume these 3 parameters are known in the 
following. Also $B \mu$ was set to be real by the field-redefinition (this is 
our convention) \cite{DU etal}, and then the soft parameter $\arg(B)$ should 
be fixed from $\arg(\mu)$ which is determined from 
$e^+ e^- \rightarrow \tilde{\chi_i}^+ \tilde{\chi_j}^-$.

In the process $e^+e^- \rightarrow \tilde{t}_i {\tilde {t_j}}^*~(i,j=1,2)$, 
there are 5 unknown parameters, $\xi$, $M_{{\tilde t}_L}^2$, 
$M_{{\tilde t}_R}^2$, $|A_t|$ and $\arg(A_t)$. However, the outgoing particles
in this case are scalar unlike the case of the chargino pair production.
Therefore only three equations are possible from the measured values of 
$\cos^2\theta_{\tilde t}$, $m_{{\tilde t}_1}^2$ and $m_{{\tilde t}_2}^2$,   
which are functions of $\xi, M_{{\tilde t}_L}^2, M_{{\tilde t}_R}^2,
|A_t|,\arg(A_t)$. ( The value of $\cos^2\theta_{\tilde t}$ will be determined 
from $\sigma_L/\sigma_R$ \cite{BE etal}.) Therefore, it is impossible to 
find the parameters of the stop sector from this process alone.
In addition, since $\xi$ depends on the Higgs sector, the process  
$e^+e^- \rightarrow \tilde{t}_i {\tilde {t_j}}^*~(i,j=1,2)$ alone can not 
provide sufficient informations to fix these five unknown parameters.
Other processes are needed. Anyway, we have three other independent
equations, {\it i.e.} the four minimum energy conditions 
($T_\phi=\partial {\cal V}_{Higgs}^{eff}/ \partial \phi =0$), one of which 
is trivial from the Goldstone's theorem \cite{{PI297},{PI373}}, that is to say,
there remain only three equations. Those equations have 9 real parameters,
$|B|$, $\xi$, $\mu_1^2$, $\mu_2^2$, $|A_t|$, $\arg(A_t)$, $|A_b|$, $\arg(A_b)$
and $M_{\rm{SUSY}}^2$ (see Ref.~\cite{PIWA}). For example, one equation 
({\it c.f.} Eq.~(\ref{Ta})) is 
\beqar
\label{Ta'}
T_a &=& \frac{\partial {\cal{V}}_{Higgs}^{eff}}{\partial a}
=0
= -v
\left[\,
{\bf Im}(m_{12}^2 e^{i \xi})
+{\bf Im}({\lambda}_5 e^{2 i \xi}) v^2 \sin\beta \cos\beta
+\frac{1}{2}{\bf Im}({\lambda}_6 e^{i \xi}) v^2 \cos^2\beta
\right.
\nonumber
\\
&&~~~~~~~~~~~~~~~~~~~~~~~~~~~
\left.
+{\bf Im}({\lambda}_7 e^{i \xi}) v^2 \sin^2\beta
\,\right]~,
\eeqar
where 
\beqar
\label{lambda5}
\lambda_5 &=&
\frac{h_t^4 \mu^2 A_t^2}{M_{\rm SUSY}^4} f_5(h_t, h_b, g_s, M_{\rm SUSY}^2,
{\bar{m_t}}^2)
+\frac{h_b^4 \mu^2 A_b^2}{M_{\rm SUSY}^4} g_5(h_t, h_b, g_s, M_{\rm SUSY}^2,
{\bar{m_t}}^2)
\\
\label{lambda6}
\lambda_6 &=&
\frac{h_t^4 |\mu|^2 \mu A_t^2}{M_{\rm SUSY}^4} f_6(h_t, h_b, g_s, 
M_{\rm SUSY}^2, {\bar{m_t}}^2)
\nonumber
\\
&&
+\frac{h_b^4 \mu}{M_{\rm SUSY}}
\left(
\frac{6A_b}{M_{\rm SUSY}}-\frac{|A_b|^2 A_b}{M_{\rm SUSY}^3} 
\right)
g_6(h_t, h_b, g_s, M_{\rm SUSY}^2, {\bar{m_t}}^2)~,
\\
\label{lambda7}
\lambda_7 &=&
\frac{h_b^4 |\mu|^2 \mu A_b^2}{M_{\rm SUSY}^4} f_7(h_t, h_b, g_s, 
M_{\rm SUSY}^2, {\bar{m_t}}^2)
\nonumber
\\
&&
+\frac{h_t^4 \mu}{M_{\rm SUSY}} 
\left( 
\frac{6A_t}{M_{\rm SUSY}}-\frac{|A_t|^2 A_t}{M_{\rm SUSY}^3}
\right)
g_7(h_t, h_b, g_s, M_{\rm SUSY}^2, {\bar{m_t}}^2)~,
\eeqar
where $f_i$ and $g_i$ ($i=5,6,7$) are of form 
\beq
f(x_1, x_2, x_3, x_4, x_5)
=k \left[
1-\frac{1}{16\pi^2} \left(
a x_1^2 + b x_2^2 + 16 x_3^2
\right) \ln{\frac{x_4}{x_5}}
\right]
\eeq
and $k$ is $O(10^{-3})$ and $a$, $b$ are $O(1)$. 
For a small ${\rm mass}^2$ splitting between the stop mass eigenstates,
$M_{\rm{SUSY}}^2=\frac{1}{2}(m_{{\tilde t}_1}^2 
+m_{{\tilde t}_2}^2)=\frac{1}{2} \rm{tr}({\cal M}_{\tilde t}^2)$  
(see Ref.~\cite{CAES etal} and the Appendix of Ref.~\cite{PIWA}),
so $M_{\rm{SUSY}}^2$ is known. Three parameters $\xi$, $|A_t|$ and $\arg(A_t)$
were already counted when we considered the processes 
$e^+e^- \rightarrow \tilde{t}_i {\tilde {t_j}}^*~(i,j=1,2)$.
Therefore, there are 11 real parameters and 7 independent equations from the
process $e^+e^- \rightarrow \tilde{t}_i {\tilde {t_j}}^*~(i,j=1,2)$ and the 
vanishing tadpole conditions. Still four parameters remain undetermined. 
However, for small $\tan\beta = O(1)$, the sbottom trilinear coupling 
can be ignored, {\it i.e.} $h_b \simeq 0$. As a result, the effects of $|A_b|$
and $\arg(A_b)$ can be effectively neglected. Thus only two parameters are 
left undetermined for low $\tan\beta$. 
At this point, let us note that the masses of two neutral Higgs bosons give 
two independent equations from Eq.~(\ref{Mh}). The masses $M_{H_i}$ of the 
light Higgs bosons can be determined from $e^+ e^- \rightarrow H_i Z^0$ 
($i=1,2$) \cite{{PIWA},{CAZE etal},{KA},{Driesen}} (since the $H_1 Z^0$ 
production gives an information about the Higgs mass $M_{H_1}$ and has a cross 
section at $\sqrt{s}=500$ GeV about $O(10)$ times larger than the 
${\tilde {t_1}} {\tilde {t_1}}^* H_1$ production (for example, see 
Ref.~\cite{Driesen}), the associated Higgs production may not give more 
information), $e^+ e^- \rightarrow H_i H_j$ ($i,j=1,2$) \cite{DE}, 
$u \bar{d} \rightarrow H_i W^+~(i=1,2)$ \cite{{PIWA},{CALY}}. 
Therefore, once the neutral Higgs masses are determined by the experiments, 
we will have two necessary informations, and finally we can determine all
the parameters. For intermediate $\tan\beta \lesssim O(20)$, $h_b$ 
can not be neglected and two parameters $| A_b |$ and $\arg A_b$ remain to 
be fixed. In such a case, the masses of two other charged and 
neutral Higgs bosons will provide the necessary informations.

Of course, we have to consider the lower bound of the lightest Higgs mass
and the EDM constraints for the above arguments.
As we have discussed above, the determinations are not very simple like
the chargino pair production \cite{CH etal} since the equations are more 
complicated.

\section{Conclusions}

We have analyzed how the CP violating phase in $A_t$ parameter 
can affect the relative phase $\xi$ of two Higgs fields, the stop masses and 
cross sections in the two processes  $e^+ e^- \rightarrow
{\tilde{t_1}} {\tilde{t_1}}^* ( H_1 )$, which may be within the scope of the 
next linear colliders. 
The relative phase ($\xi$) of the two Higgs fields that arise from the quantum 
corrections has very small dependence on $\arg(A_t)$ and $|\mu|$. The sizes of 
$\xi-\pi(\rm and ~2\pi)$ are also numerically very small. 
In the process $e^+ e^- \rightarrow {\tilde{t_1}} {\tilde{t_1}}^*$, the stop 
mixing angle $\theta_{\tilde t}$, the stop masses $m_{{\tilde t}_i}$ ($i=1,2$)
and the cross sections have very little dependence on $\xi$, but 
strong dependence on $\arg(A_t)$ since the stop mass itself changes a lot
when $\arg(A_t)$ varies. The typical cross section is order of a several tens
of fb, which is well within the scope of NLC's.
In the process $e^+ e^- \rightarrow {\tilde{t_1}} {\tilde{t_1}}^* H_1$, 
the loop correction becomes very important, since the lightest Higgs boson 
mass can be strongly affected by loop corrections with complex trilinear 
coupling $A_t$. However the neutral
Higgs boson mixing matrix or $\xi$ does not change significantly even if 
quantum corrections are included. In this case, the typical cross section is 
order of $O(0.1 - 10)$ fb depending on the $A_t$ phase and $| \mu |$.
Therefore this mode can be (less) promising at NLC's depending on the 
soft SUSY breaking parameters.  

Since important physical quantities (stop mass and cross sections) may depend
on the CP phases of soft SUSY breaking parameters, it is necessary to have a
strategy for determination of these parameters in the presence of CP violating
phase. We argued that, for low $\tan\beta$ region, it is in fact possible 
to fix $A_t$ and other mass parameters in the Higgs and stop sectors through 
$e^+ e^- \rightarrow \tilde{t}_i {\tilde{t_j}}^*$, the vanishing tadpole 
conditions and the masses of neutral Higgs bosons. For intermediate 
$\tan\beta$ upto $\sim 20$, the situation is the same if one has informations 
on the masses of two other charged and neutral Higgs bosons.    
The whole procedure to determine soft parameters seems complicated,
but it will be possible.

\section*{Acknowledgements}
The author thanks Professor Pyungwon Ko for useful ideas and 
suggestions and Jae Hyun Park for helps in drawing figures.
This work is supported by the Brain Korea 21 Project and grant No. 
1999-2-111-002-5 from the
interdisciplinary Research program of the KOSEF.
\newpage

\newpage
\begin{figure}
\begin{center}
\setlength{\unitlength}{0.240900pt}
\ifx\plotpoint\undefined\newsavebox{\plotpoint}\fi
\sbox{\plotpoint}{\rule[-0.200pt]{0.400pt}{0.400pt}}%
%
\end{center}
\caption{\label{dstpm5}The variations of the total cross sections 
with respect to
the $\arg(A_t)$ in the process $e^+ e^- \rightarrow {\tilde{t_1}}
{\tilde{t_1}}^* $~ ($\sqrt{s}=500$ GeV, $\mu= 500$ GeV, $|A_t|=250$ GeV,
$M_a \approx 194$ GeV,
$\tan\beta=2$ and $M_{\rm SUSY}=500$ GeV\,)}  
\end{figure}
\begin{figure}
\begin{center}
\setlength{\unitlength}{0.240900pt}
\ifx\plotpoint\undefined\newsavebox{\plotpoint}\fi
\sbox{\plotpoint}{\rule[-0.200pt]{0.400pt}{0.400pt}}%
%
\end{center}
\caption{\label{dcm5} The variations of the total cross sections 
with respect to 
$\arg(A_t)$ in the process $e^+ e^- \rightarrow {\tilde{t_1}} 
{\tilde{t_1}}^* H_1$~ ($\sqrt{s}=500$ GeV, $\mu= 500$ GeV, $|A_t|=250$ GeV, 
$M_a \approx 194$
GeV, $\tan\beta=2$ and $M_{\rm SUSY}=500$ GeV\,)} 
\end{figure}
\begin{figure}
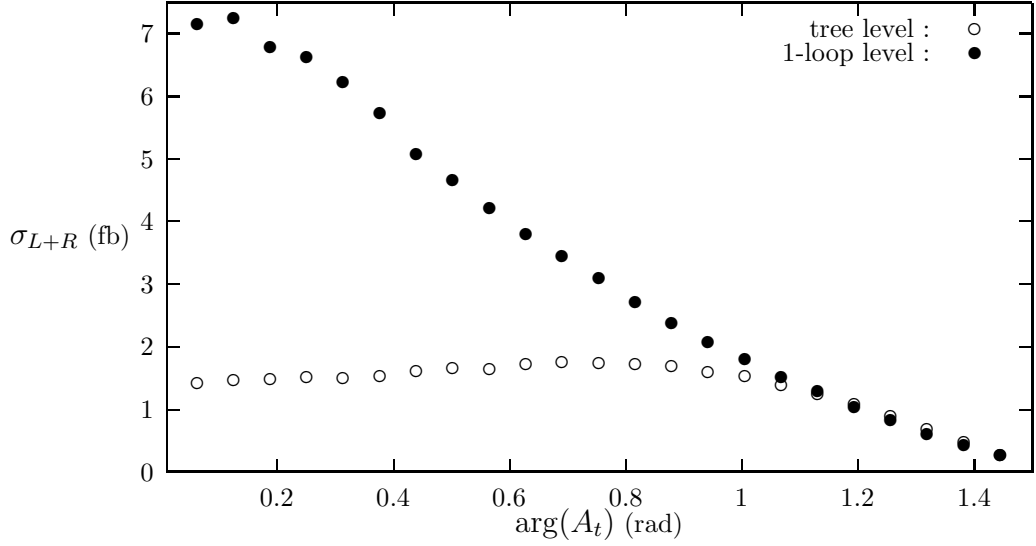

\begin{center}
\setlength{\unitlength}{0.240900pt}
\ifx\plotpoint\undefined\newsavebox{\plotpoint}\fi
%
\end{center}
\caption{\label{dcm10} The variations of the total cross sections
with respect to
$\arg(A_t)$ in the process $e^+ e^- \rightarrow {\tilde{t_1}}
{\tilde{t_1}}^* H_1$~ ($\sqrt{s}=500$ GeV, $\mu= 1000$ GeV, $|A_t|=500$ GeV,
$M_a \approx 
274$ GeV, $\tan\beta=2$ and $M_{\rm SUSY}=500$ GeV\,) Note that the range 
of $\arg(A_t)$ is [0, 1.5] approximately, because the lightest Higgs mass
$M_{H_1}$ is imaginary elsewhere owing to the large $|\mu|$.}
\end{figure}
\end{document}